# Analysis of Cargo Loading Modes and Capacity of an Electrically-Powered Active Carrier

Xiaoye Huo, Yue Wu, Alicia Boymelgreen and Gilad Yossifon*

Faculty of Mechanical Engineering, Micro- and Nanofluidics Laboratory, Technion – Israel Institute of Technology, Haifa 32000, Israel

* Corresponding author: yossifon@technion.ac.il

## Abstract

The use of active colloids for cargo transport offers unique potential for applications ranging from targeted drug delivery to lab-on-a-chip systems. Previously, Janus particles (JPs), acting as mobile microelectrodes have been shown to transport cargo which is trapped by a dielectrophoretic mechanism [Boymelgreen et al. (2018)]. Herein, we aim to characterize the cargo loading properties of mobile Janus carriers, across a broad range of frequencies and voltages. In expanding the frequency range of the carrier, we are able to compare the influence of different modes of carrier transport on the loading capacity as well as highlight the differences between cargo trapped by positive and negative dielectrophoresis. Specifically it is shown that cargo trapping results in a reduction in carrier velocities with this effect more pronounced at low frequencies where cargo is trapped close to the substrate. Interestingly, we observe the existence of a maximum cargo loading capacity which decreases at large voltages suggesting a strong interplay between trapping and hydrodynamic shear. Finally, we demonstrate that control of the frequency can enable different assemblies of binary colloidal solutions on the JP. The resultant findings enable the optimization of electrokinetic cargo transport and its selective application to a broad range of targets.

# 1. INTRODUCTION

The use of active colloids for microscale cargo transport has the potential to impact a broad range of research areas with applications including drug delivery[1], detoxification[2], environmental remediation[3], immunosensing[4] and even SMART material repair[5]. As opposed to phoretically driven transport, which is characterized by mass migration in response to externally imposed gradients, active particles asymmetrically draw and dissipate energy at the colloidal scale, creating local gradients which drive autonomous propulsion[6]. Since the driving force is produced on the particle level, active colloids are free to travel along individual pathlines and can cover larger areas and volumes while operating under simpler ambient conditions (i.e., without the necessity for field or chemical gradients).

The electrokinetically driven metallodielectric Janus particles considered herein, represent a unique subset of active colloids, where the initial energy source is an externally applied field (here electric), but since the conversion to a kinetic energy occurs on the particle level, the colloids remain "active" as is evident by their individual and group behavior[7]. This subset (which also includes driving mechanisms such as self-thermophoresis[8]) maintains the advantages of colloidal scale propulsion, is fuel free – thereby avoiding issues of finite life and/or non-bio-compatibility of commonly used fuels, such as hydrogen peroxide[9][10] and perhaps most importantly, offers the ability to externally control parameters such as speed and direction in real time. In the specific case of electrokinetically driven metallodielectric Janus spheres, variation of the frequency of applied electric field has been shown to alter both the speed and direction[7,11], as the particles transition from translating with dielectric hemisphere forward under ICEP[12] to moving with the metallic hemisphere forward under sDEP[11].

Recently, we have demonstrated[13] that as well as driving particle motion, the interaction of the Janus sphere with the electric field also yields localized electric field gradients which can be used to manipulate cargo via dielectrophoresis[14], wherein a particle will either be attracted to (positive DEP, pDEP) or repelled from (negative DEP, nDEP) regions of high field strength according to its geometry and material properties relative to the suspending electrolyte. The significance of this finding is that dielectrophoresis offers a label-free method to selectively and dynamically manipulate (load, transport and release) a broad range of organic and inorganic cargo.

Combining DEP with electrically powered active particle propulsion yields an active carrier that can selectively load, transport and release a broad range of cargos, singularly controlled by an external electric field. This unification allows for significantly simpler and more robust operation when compared to the more traditional approach to cargo transport wherein propulsion of the active carrier and cargo manipulation (load and release) were often considered as disparate problems, realized by combining two different mechanisms; self-propulsion can be driven by e.g., electric[15], magnetic[16] and optical[17] external fields[18] and even with chemical fuel[19], while cargo loading is achieved by e.g., magnetic[20], electrostatic[21], biomolecular[22] recognition and attraction mechanisms.

In order to advance the application of this unique cargo transport mechanism, in this work we aim to characterize the interplay between cargo loading and transport and highlight necessary parameters for the system optimization. The addition of magnetic steering, e.g., magnetizing a ferromagnetic Ni layer coated on half of the Janus particle surface[23], enables directed motion via an external rotating static magnet and the quantification of maximal cargo loading. In contrast, to previous studies of electrically driven cargo loading, which were either limited to stagnant JPs[13] or limited to high frequencies[23] where electro-convection effects diminish, we study the cargo loading dependency of mobile JP across a broad spectrum of frequnecy (kHz-Mhz) and voltage. As a result, we are able to characterize trapping for both pDEP and nDEP conditions as well as the effect of the propulsive mechanism (ICEP or sDEP) on the cargo loading.

## 2. EXPERIMENTAL MATERIALS AND METHODS

*Magnetic Janus particle fabrication:* Polystyrene particles (diameter: 10μm) (Sigma Aldrich) in isopropanol (IPA) were pipetted onto a glass microscope slide, to form a monolayer of particles upon solvent evaporation. The glass slide was coated with 15nm Cr, followed by 50nm Ni and 15nm Au, as described in the protocol outlined in Ref.[24] and Ref.[25]. To magnetize the Janus particles, the substrates were placed in between two neodymium magnetic blocks (14x12x19mm in size), with opposite dipoles facing each other. Next, the substrate was sonicated in Distilled water (2% Tween 20 (Sigma Aldrich)) to release the Janus particles. The Janus particles were then washed three times in DI water (0.01%Tween 20(Sigma Aldrich)) before the experiment.

*Magnetic steering of Janus particles:* Guiding the direction of the moving Janus particle was realized by placing the neodymium magnet block (14x12x19mm in size) at a specific orientation close to the microchamber. The magnet was kept at a horizontal distance of 3cm from the focus of the objective and at the same height as the microchamber. In this setup, the magnet produces a field of 125 Gauss in the microchamber.

*Polystyrene tracer particles solution preparation:* For observation of the 10 µm JP and 1-3 µm cargo particles (Polystyrene particles, Fluoro-max), particles were rinsed three times with DI water, to which 0.01% of nonionic surfactant (Tween 20 (Sigma Aldrich)) was added in order to minimize adhesion to the ITO substrate before being injected into the microfluidic chamber via a small hole at the upper substrate, drilled expressly for this purpose. The concentration of the PS particle is 0.01-0.05% (w/v).

*Experimental set-up:* The experimental chamber consisted of a 120µm-high, silicone reservoir (Grace-Bio), sandwiched between an ITO-coated, 1mm glass slide (Delta Technologies) and an ITO-coated coverslip (SPI systems), as illustrated in Ref [11]. The bottom ITO-coated coverslip was further coated with 15nm silicon dioxide using sputtering deposition (AJA International Inc., ATC 2200) to prevent adsorption of the particle onto the substrate[7]. Two inlet holes (~1mm in diameter) were drilled through the top 1mm ITO slide, surrounded by a silicone reservoir (2mm in height and 9mm in diameter) filled with solution, to ensure the chamber remained wet and to enable the addition of the solution with the JPs and cargo particles into the channel via manual pumping. The AC electrical forcing was applied using a signal generator (Agilent 33250A) and monitored by an oscilloscope (Tektronix-TPS-2024). A lab-made switch (Solid State Relays (AQV252G) controlled by Arduino processer) was used to control the duration and timing of AC pulses.

*Microscopy and image analysis:* The trapping behavior of the particles was observed using a Nikon Eclipse Ti-E inverted microscope equipped with a Yokagawa CSU-X1 spinning disk confocal scanner and Andor iXon-897 EMCCD camera. The chamber was placed with the coverslip side down and images were taken using a ×60 oil immersion lens. The velocity of the JP and trapped cargo was processed and calculated from the image sequences recorded with an inverted epi-fluorescent microscope (Nikon, Eclipse Ti-U) using a ×20 objective.

*Numerical simulations:* The numerical simulation used to qualitatively verify the presence of asymmetric electric field gradients arising from the proximity of a Janus sphere near a conducting wall, was performed in COMSOL™ 5.3. A simple 2D geometry, consisting of a rectangular channel, 50 µm height and 50 µm width, with a 10 µm diameter circle placed 200nm above the substrate, was used to model the experimental setup[13]. Since the EDLs are thin relative to the radius of the particle $(\lambda/a \ll$ , within the electrolyte we can solve the Laplace equation for the electric potential, $\phi$, in conjunction with the following boundary condition at the metallic side of the JP $\sigma \frac{\partial \phi}{\partial n} = i\omega C_{DL}\left(\phi - V_{floating}\right)$, which describes the oscillatory Ohmic charging of the induced EDL, wherein $V_{floating}$ is the floating potential of the metallic hemisphere of the JP, $n$ is the coordinate in the direction of the normal to the JP surface, and $C_{DL}$ represents the capacitance per unit area of the EDL and may be estimated from the Debye-Huckel theory as $C_{DL} \sim$    . In addition, a floating boundary condition[12] was applied on the metallic hemisphere (right) so as to obey total zero charge. An insulation boundary condition was applied on the dielectric hemisphere (left) of the JP, a voltage of 6.25 V was applied at the lower substrate ($y$=0), while the upper wall was grounded, and the edges of the channel were given an insulating boundary condition.

## 3. RESULTS AND DISCUSSION

Magnetic metallodielectric Janus spheres, 10μm in diameter are placed in the experimental setup depicted in Fig.1. In conjunction to the parallel indium tin oxide (ITO) coated glass slides that induce the electric field which drives JP propulsion and cargo manipulation, we have also incorporated an option for magnetic steering via a motorized static magnet. This steering capability enables directed motion for the sequential pickup of cargo in solutions with low concentration of cargo, enabling precise quantification of the maximum load the carrier can bear (see Video 1 in SI). The cargo is trapped by a dielectrophoretic mechanism[14] wherein it is either attracted to (pDEP) or repelled from (nDEP) regions of high electric field strength according to its geometric properties and the frequency dependent, relative material properties of the target and solution. In Fig. 1c, we plot the numerical simulation of the localized electric field around the JP, noting several distinct locations of dielectrophoretic potential wells for either pDEP (locations 1 and 2) or nDEP (locations 3 and 4) corresponding with regions of high and low electric field intensity, respectively.

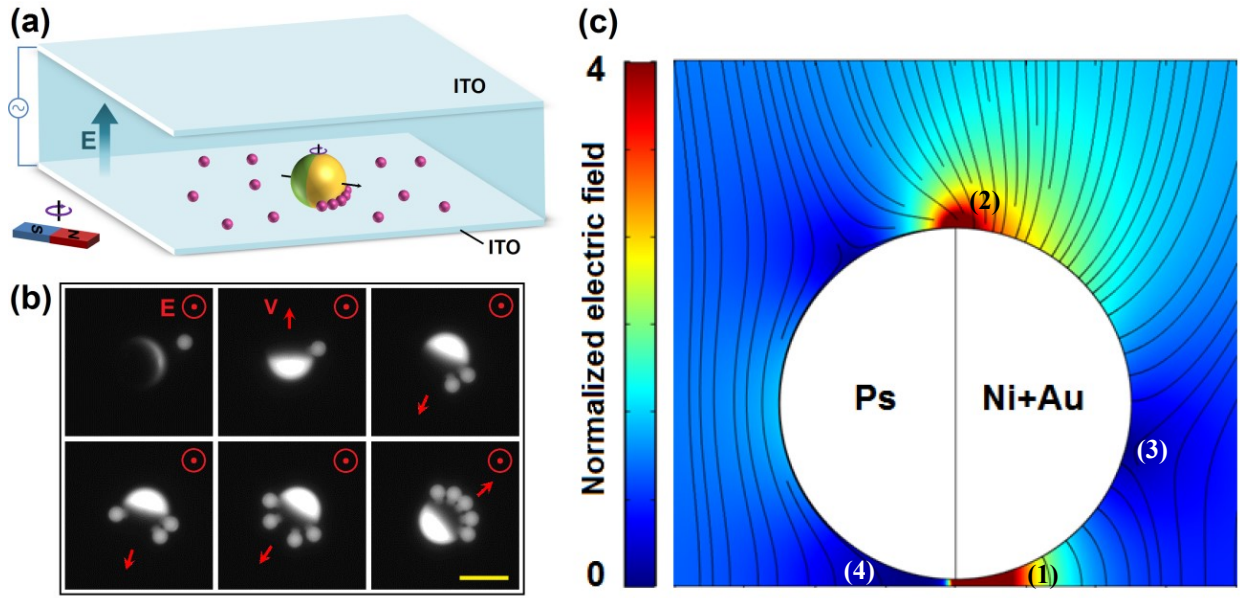

**Fig.1: Experimental setup and methodology.** (a) Schematics of the microchamber including the JP carrier and the trapped 2µm cargo particles. The magnetic field is used for steering of the JP via the ferromagnetic Ni coating while the electric field is used for both propulsion and cargo manipulation; (b) Using magnetic steering for individual pickup of 3µm cargo in order to study the velocity of the carrier as a function of the number of cargo (see Video 1 in SI). AC voltage of 5V and 1.5MHz is applied. The direction of the JP velocity is indicated with red arrows. The scale bar is 10µm; (c) Numerical simulation results showing the electric field distribution around the JP that is in vicinity to the powered electrode and the corresponding locations where positive (locations 1 and 2) and negative (locations 3 and 4) DEP trapping of cargo can occur.

In order to fully characterize the potential of this cargo transport system and evaluate the optimal conditions for transport of a given cargo, it is necessary to evaluate the interplay between the various physical parameters governing both the carrier motion and cargo loading mechanism. In Figure 2 (see also Video 2 in SI), we highlight the influence of the frequency of the electric field, which controls both the carrier's mode of transport (ICEP or sDEP) and the dielectrophoretic response of the target. At low frequencies, where the carrier is propelled in the direction of its dielectric end due to induced-charge electrophoresis[12], the aft positioning of the cargo and strong hydrodynamic flow around the metallic hemisphere (which is the superposition of the quadrupolar flow due to ICEO and the motion of the particle), causes the targets undergoing pDEP and trapped aft of the JP to concentrate around the x-axis (Fig.2b).

In contrast, at high frequencies, where the carrier travels with the metallic hemisphere forward, the target, now facing fore, is more uniformly distributed around the circumference of the metallic hemisphere (Fig.2c). As the frequency of the external field is increased, the response of the target transitions from pDEP to nDEP, (Fig.2c), resulting in their repulsion from the region of maximum electric field strength (location 1 in Fig.1c). Instead, targets are trapped at the local minima and can be observed at both the metallic equator of the JP (location 3 in Fig.1c) and underneath the dielectric hemisphere (location 4 in Fig.1c) (Fig. 2d). It is emphasized that the ability to trap targets in low potential wells under nDEP conditions is of immense practical importance for the application of this transport system to biological matter such as cells which are expected to exhibit nDEP in high conductivity physiological media[26].

In the case of targets undergoing pDEP, geometric considerations dictate that the minimum distance of the trapped cargo from the center of the JP ($x_1$) will increase with increasing cargo size ( $x_1 = 2\sqrt{a_{JP} \cdot a_{cargo}}$ where $a_{JP}$ and $a_{cargo}$ are the radii of the JP and trapped target, respectively[13]). Since the intensity of the electric field – and correspondingly the DEP force – decreases rapidly with distance from the JP[13], there exists a critical applied voltage, at which the electric field at $x_1$ is sufficient to trap the target. For a stagnant Janus sphere, an expression for this critical voltage is obtained in [13] by balancing the thermal and dielectrophoretic potentials acting on the target and is given by $V_{RMS}^2 \sim 0.3(1 + 7 x_1/a)$. For the current conditions of $2a_{JP} = 10\mu m$ and $2a_{cargo} = 3\mu m$, the critical voltage is equal to $V_{RMS} \sim 1.6V$ or $4.6V_{p-p}$. Noting that in the current work, the carrier is mobile and thus a force balance must also account for Stokes drag which acts against dielectrophoretic trapping, we expect that the critical voltage for a given target size will be higher than that given in[13]. This difference is evident in Figure 3b where we observe that the JP cannot trap the cargo underneath the metallic hemisphere (location 1 in Fig.1c) under ICEP propulsion and pDEP conditions. Interestingly, under reversed motion of the carrier, the larger targets are trapped at the top of the Janus sphere (Fig.3c). The reason that no such trapping on the top of the JP is observed in the case of 2μm (Fig.2) might be simply due to the lower pDEP that these particles undergo relative to that of the 3μm, and hence, are hydrodynamically sheared.

In Fig.2e,f and Fig.3e,f, we illustrate the effect of target accumulation on the velocity of the cargo carrier. Overall, as the number of accumulated targets increase, the velocity of the carrier monotonically decreases. At high frequencies, where the JP is undergoing sDEP propulsion while the cargo undergo nDEP trapping, this retardation can be approximated by accounting for the increased Stokes drag arising from increased surface area in agreement with the result of Demirörs [26]

$$v_{nDEP}(i) = F_{Propulsion} \big/ 6\pi\eta a_{eff} = F_{Propulsion} \big/ 6\pi\eta \sqrt{a_{JP}^2 + i \cdot a_{cargo}^2} \; , \quad (1)$$

where, the effective radius $a_{eff}$ is a superposition of the Janus particle radius, $a_{JP}$, and that of the cargo particle, $a_{cargo}$, where $i$ is the number of the trapped cargo. The only fitting parameter is the propulsion force that drives the particle, $F_{Propulsion}$ (sDEP), which is assumed be constant, for a specific frequnecy and applied voltage, irrespective of the number of trapped cargo. This force is shown to scale quadratically with the applied voltage (see Fig.S2 in SI) in agreemnt with Ref.[13].

At low frequencies however, where the Janus sphere is propelled by ICEP, this simplified model significanlty overpredicts the velocity of the carrier - suggesting that additional physics are at play. One potential cause for the extra reduction in velocity measured in the experiments could be the disruption of the ICEO flow due to the target whose presence both disrupts the EDL formed at the metallic hemisphere, impacting the formation of the ICEO flow in a manner similar to a slip-stick surface (e.g.[27]) as well as constituting a physical barrier to the induced-charge electroosmotic flow which propels the carrier. However, examination of this possibility in both 2D and 3D numerical simulations in COMSOL at a frequency corresponding to the RC time of the induced charge (see Fig.S3 in SI) and a single trapped cargo particle indicates velocity reductions of only ~8% and ~1%, respectively - significantly lower than the experimental observations of ~7, ~13 and ~56% for the 1, 1.5 and 3kHz cases respectively. Note that the 2D case is an overstimation as the actual ratio of the cargo cross-section relative to that of the JP is much smaller in the realistic 3D case.

Noting that at low frequencies, the targets are trapped by pDEP closed to the substrate (rather than at the equator as is the case for high frequency nDEP), an alternative possibility for this discrepency is near wall effects; due to either increased Stokes drag on the cargo adjacent to the wall[28] and/or electrostatic interaction of the cargo with its image dipole[29]. In line with this theory, in Figure 2f we have successfully used a single constant cargo-wall interaction force, $F_{wall}$, per single trapped cargo to fit the three cases of a carrier moving under ICEP, such that

$$v_{pDEP}(i) = \left(F_{Propulsion} - i \cdot F_{wall}\right) \big/ 6\pi\eta\sqrt{a_{JP}^2 + i \cdot a_{cargo}^2} \; . \quad (2)$$

where the significant reduction with increased frequency is explained by noting that as the velocity of the carrier decreases, the relative impact of cargo-wall interaction on the mobility increases. For the of 2µm cargo case depicted in Fig.2d we obtained $F_{wall} \sim 0.19\,pN$ and $F_{Propulsion} \sim 3.13\,pN$.

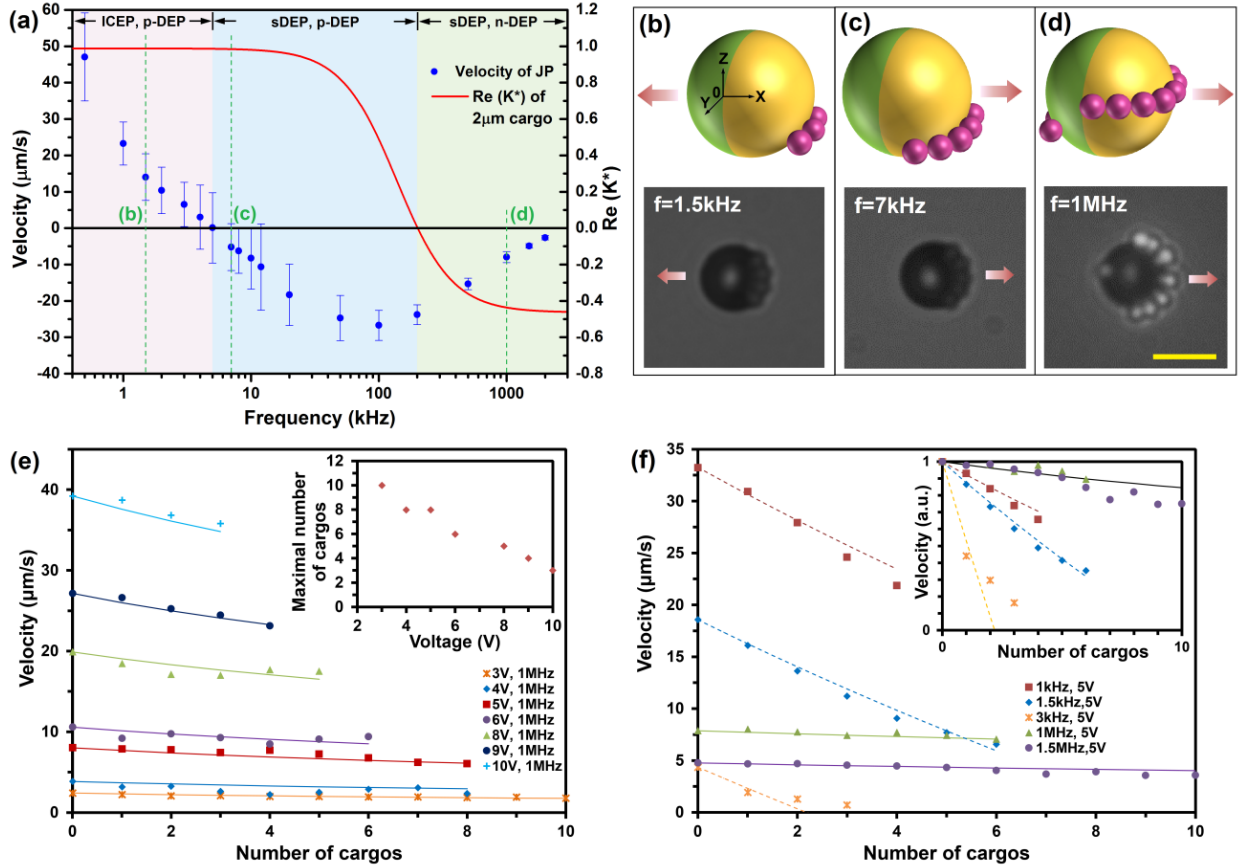

**Fig.2: Cargo transport based on both positive dielectrophoresis (pDEP) and negative dielectrophoresis (nDEP) trapping of 2µm cargo.** (a) For the same cargo size of 2µm we demonstrate transport of both pDEP and nDEP trapped cargo depending on the applied electric field frequency. Three frequency domains are observed – each of which correspond with schematics and microscope images b-d (see also Video 2 in SI). (b) At low frequencies, the cargo particles undergo pDEP and Janus particle translates forward under induced-charge-electro-phoresis (ICEP). (c) Janus reverses direction under self-dielectrophoresis (sDEP) while cargos are still pDEP trapped. (d) nDEP trapped cargos assemble on the equator of the Janus particle. The scale bar in (b)-(d) is 10µm. The red arrows indicate the direction of the JP velocity; (e) carrier velocity as a function of the number of cargo particles for varying voltage at 1MHz. Inset depicts the maximum cargo loading versus the applied voltage. (f) carrier velocity versus applied frequency and fixed voltage. The continuous lines in parts (e) and (f) are the fitted Stokes drag model,

Eq.(1), accounting for the overall increased cross-sectional area due to the trapped cargo, while the dashed lines are the added cargo-wall interaction force of the cargo with the substrate, Eq. (2).

The maximal load of the carrier is determined by observing conditions at which the carrier approaching a target but having reached its maximal load is either unable to pick up the additional cargo, or if while picking it up, one of the previously attached targets is released. An important design parameter for cargo transport, it would seem that this maximum number of cargo reflects the interplay between hydrodynamic shear and the dielectrophoretic trapping force. Interestingly, we observe that at a frequency of 1MHz, where the carrier undergoes sDEP propulsion and cargo is trapped under nDEP at the Janus midplane, the applied voltage inversely affects the maximum number of cargo that may be trapped at the carrier surface (see insets of Fig. 2e and Fig. 3e for 2µm and 3µm cargo particles, respectively). Specifically, as the voltage and therefore carrier velocity increases, the maximal number of cargo that can be trapped decreases (see inset of Fig.2e of 2µm cargo), suggesting that at high voltages the increased trapping force (which scales with $E^2$) is outweighed by an increased hydrodynamic shear force. It would seem that this increased shear is not exclusively due to the increased carrier velocity (which is also expected to scale with $E^2$ (see Fig.S2 in SI) and therefore maintain the same balance with the DEP force), but may also be related to the particle-particle interactions of the targets. This proposal is supported by the videos [e.g. Video 3 in SI] where the cargo being released appears to be pushed/bumped from the interface by the adjacent target. A more detailed analysis of the forces acting on the trapped cargo is left for future work.

This same trend is also observed for the 3µm cargo particles (inset of Fig.3e) at high voltages. At low voltages however, we observe that the trapping increases with the applied field – likely due to the emerging dominance of dielectrophoretic over thermal potential[13] suggesting an optimal voltage at which a maximum number of targets is trapped. It is posited that this maxima is a general characteristic of the system and that for sufficiently small voltages, a similar maxima would be observed for 2µm cargo but this could not be evaluated in the current experimental setup as at voltages below 3V the JP velocity was too small to enable stable propulsion without temporary stops. Attempts at obtaining similar analysis in the lower frequency regime of ICEP propulsion also did not yield reliable results as the carrier tended to get stuck due to the increased interaction of the loaded cargo with the substrate. Once stuck, the concentration of cargo particles is diffusion limited due to the diluted concentration.

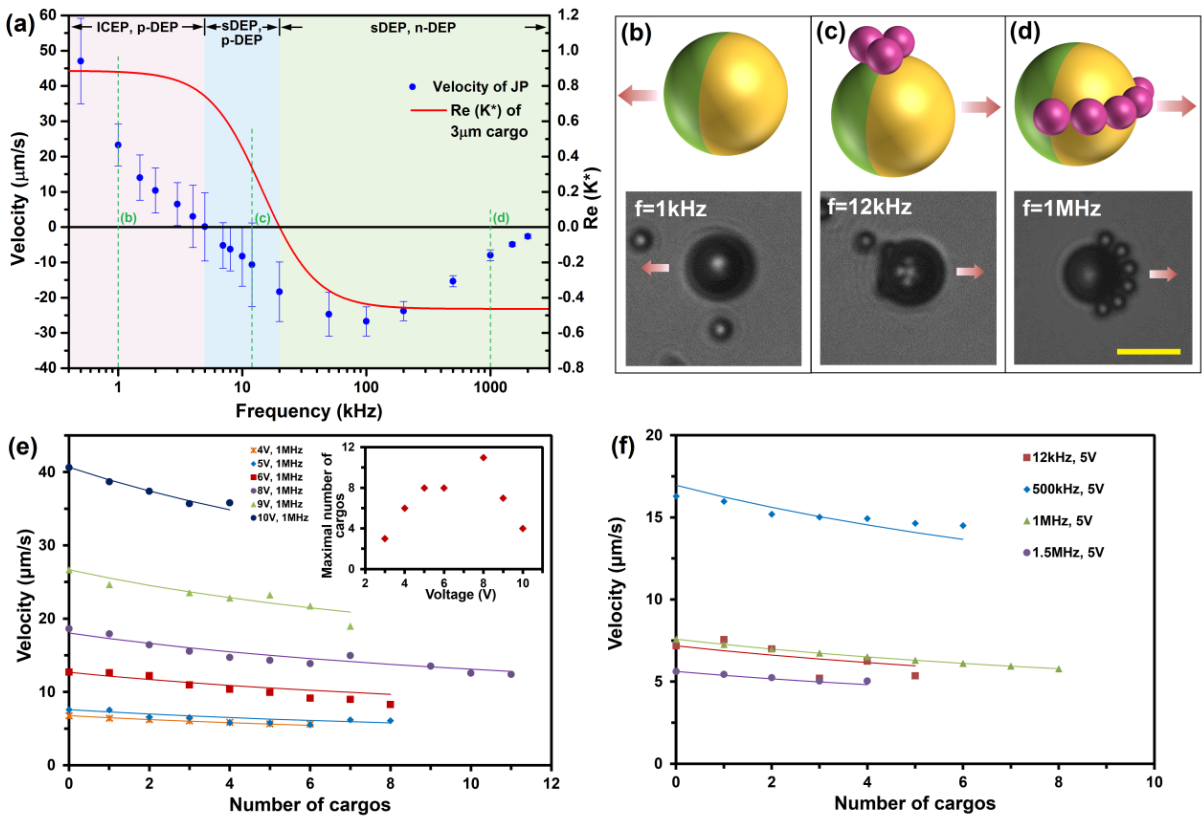

**Fig.3: Cargo transport based on both positive dielectrophoresis (pDEP) and negative dielectrophoresis (nDEP) trapping of 3μm cargo.** (a) For the same cargo size of 3μm we demonstrate transport of both pDEP and nDEP trapped cargo depending on the applied electric field frequency. Three frequency domains are observed – each of which correspond with schematics and microscope images b-d (see also Video 3 in SI). (b) At low frequencies, the cargo particles undergo pDEP but are not being able to be trapped under the metallic hemisphere. The Janus particle translates forward under induced-charge-electro-phoresis (ICEP). (c) Janus reverses direction under self-dielectrophoresis (sDEP) while cargos are pDEP trapped on top of the JP. (d) nDEP trapped cargos assemble on the equator of the Janus particle. The scale bar in (b)-(d) is 10μm. The red arrows indicate the direction of the JP velocity; (e) carrier velocity as a function of the number of cargo particles for varying voltage at 1MHz. Inset depicts the maximum cargo loading versus the applied voltage; (f) carrier velocity versus applied frequency and fixed voltage. The continuous lines in parts (e) and (f) are the fitted Stokes drag model accounting for the overall increased cross-sectional area due to the trapped cargo.
11

One of the key properties which makes DEP based cargo trapping so attractive is the selective nature of the force, where the unique geometry of the target and its material properties relative to the solution determine whether the cargo will be attracted or repelled from regions of high field strength. In the following section, we examine cargo manipulation of a binary particle solution and it is demonstrated that we can either trap a mixture or separate it at the surface of the JP according to the frequency of the applied field (Fig.4; see also Video 4 in SI). For an applied voltage of 5V, at frequencies where both cargo particles exhibit pDEP, the small particles (1µm) and the larger particles (3µm) are trapped at different locations where the former is trapped between the JP and the bottom substrate while the latter is pDEP trapped only at the top of the JP (location 2 in Fig.1c) due to its inability to penetrate underneath the JP to the region of maximum electric field intensity (location 1 in Fig.1c). As the frequency is increased beyond the COF of the 3µm, the smaller particles that exhibit pDEP are still trapped between the JP and the substrate while the larger particles that exhibit nDEP are now trapped at the equator of the metallic hemisphere. Further increase of the frequency beyond the COF of the second cargo - so that both types of cargo now exhibit nDEP – results in both being trapped at the equator of the metallic hemisphere. The ability to simultaneously trap both nDEP and pDEP cargoes simultaneously at distinctly different locations on the active carrier itself, opens up new possibilities of cargo loading and transport of complex multi-particle solutions.

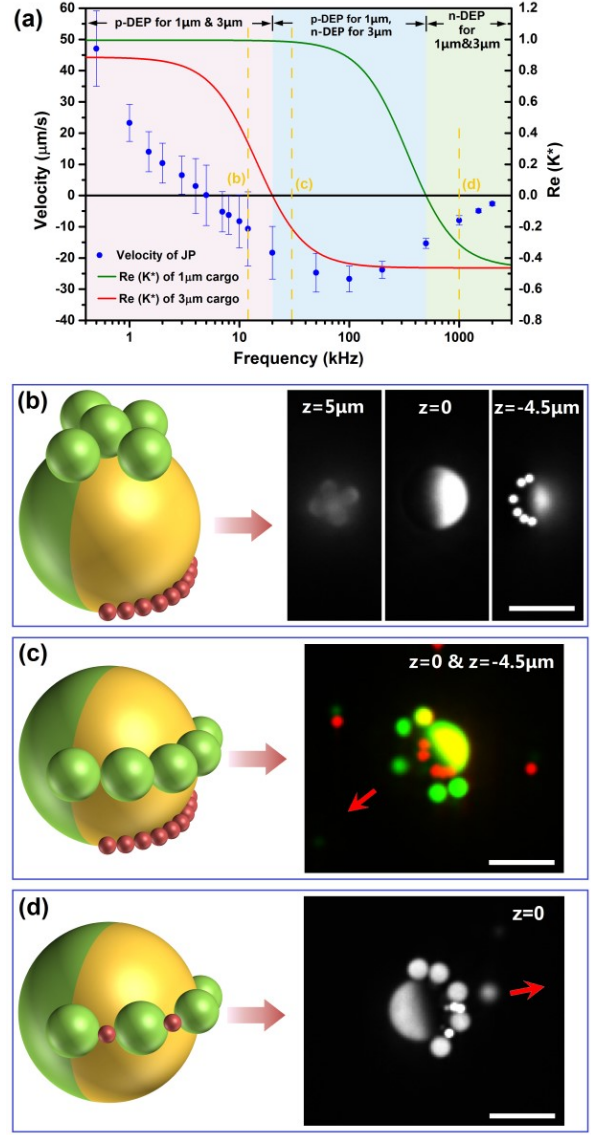

**Fig.4: Multi-cargo trapping at different JP locations.** (a) Bare carrier velocity versus frequency and corresponding CM factor of the 1 and 3μm cargo particles; (b) At 12 kHz, both 1μm and 3μm particles undergo p-DEP, and the JP trapped 3μm cargo on the top and 1μm on the bottom. Microscopy images at focal plane z=5μm (top), z=0 (equator) and z=-4.5μm (bottom) are shown. (c) At 30 kHz, when 1μm particles undergo p-DEP while 3μm particles undergo n-DEP, the JP trapped 3μm cargo on the equator and 1μm on the bottom. Two microcopy images at focal plane at z=0 and z=-4.5μm are stacked to the same plane to show the different location of cargos. (d) At 1 MHz, both 1μm and 3μm particles undergo n-DEP, and they are both trapped at the equator of the JP. See corresponding Video 4 in SI. The scale bar in (b)-(d) is 10μm. The red arrows indicate the direction of the JP velocity.

## 4. CONCLUSIONS

To summarize, we examined the interplay between different DEP response of cargo particles and the JP propulsion mechanism as controlled by the electric field frequency. Accumulation of cargo was shown to decrease the velocity of the carrier, with the reduction most marked at low frequencies where the particle is driven by ICEP. Distinct trapping positions on the JP surface were obtained depending on both frequency and particle size. For example, while small cargo particles (1µm in Fig.S4 and 2µm in Fig.2) tend to be trapped under pDEP conditions underneath the metallic coated hemisphere, larger particles (3µm in Fig.3) can only be pDEP trapped on the top of the JP as their size restricts them from being able to sample the large field gradients between the JP and the substrate. This dependency on the frequency has been also shown to enable different assemblies of a binary colloidal solution onto a JP (Fig.4). Furthermore, instead of the monotonic increased cargo loading previously observed for stagnant JP, our mobile JP exhibited maximal loading capacity at some critical voltage which decreased with further increased voltage.

## ASSOCIATED CONTENT

### Supporting Information

Additional details, figures, and videos as described in the text. The Supporting Information is available free of charge on the ACS Publications website at DOI:

## AUTHOR INFORMATION

### Corresponding Author

* E-mail: yossifon@techinon.ac.il

### Notes

The authors declare no competing financial interest.

## ACKNOWLEDGEMENTS

We wish to acknowledge the Technion Russel-Berrie Nanotechnology Institute (RBNI) and the Technion Micro-Nano Fabrication Unit (MNFU) for their technical support. G.Y. acknowledge the support from ISF Grant 1938/16, X.H acknowledges the support from Aly Kaufman fellowship and Y.W. acknowledges the support from the Technion-Guangdong fellowship.